\documentclass[preprint,12pt]{elsarticle}
\biboptions{sort&compress}
\usepackage[utf8]{inputenc}
\usepackage{comment}
\usepackage{amssymb}
\usepackage{graphicx} 
\usepackage{caption}
\usepackage{tabularx}
\usepackage{subcaption}
\usepackage{float}
\usepackage{ragged2e}
\usepackage{enumitem}
\setlist[enumerate]{after={\bigskip}}
\usepackage[utf8]{inputenc}
\usepackage[english]{babel}
\usepackage{amsmath}
\usepackage{amsfonts}
\usepackage{adjustbox}
\usepackage{amssymb}
\usepackage{graphicx}
\usepackage{float}
\usepackage{ulem}
\usepackage{miller}
\usepackage{hyperref}
\usepackage{xcolor}
\usepackage{natbib}
\usepackage{lineno}
\usepackage{xurl}
\usepackage{enumitem}
\usepackage{siunitx}
\usepackage{booktabs} 
\usepackage[version=4]{mhchem}
\usepackage[margin=2cm]{geometry}
\usepackage{lmodern} 
\usepackage{tablefootnote}
\usepackage{amsmath,amssymb}

\normalsize
\DeclareSIUnit{\angstrom}{\AA}

\usepackage{xstring}
\makeatletter
\newcommand\zoneaxis[1]{%
    \begingroup
        \expandarg
        \def\remain@arg{#1 }%
        \ensuremath{[\zoneaxis@]}%
    \endgroup}

\newcommand\zoneaxis@{%
    \StrBefore\remain@arg\space[\current@arg]%
    \StrBehind\remain@arg\space[\remain@arg]%
    \IfBeginWith\current@arg-%
        {\overline{\expandafter\@gobble\current@arg}}%
        \current@arg
    \ifx\@empty\remain@arg\else
        \thinspace\expandafter\zoneaxis@
    \fi}
\makeatother

\begin{document}
\begin{frontmatter}

\title{Revealing 3D orientation and strain heterogeneity in calcite generated by bio-cementation}

\author[inst1]{Marilyn Sarkis\corref{cor1}}
\author[inst1]{James A. D. Ball}
\author[inst2]{Michela La Bella}
\author[inst3]{Antoine Naillon}
\author[inst3]{Christian Geindreau}
\author[inst3]{Fabrice Emeriault}
\author[inst1]{Carsten Detlefs}
\author[inst1]{Can Yildirim}

\cortext[cor1]{Corresponding author: marilyn.sarkis@esrf.fr}


\affiliation[inst1]{organization={ESRF, The European Synchrotron},
                    addressline={71 Avenue des Martyrs},
                    city={38000 Grenoble},
                    country={France}}

\affiliation[inst2]{organization={Department of Physics, Technical University of Denmark},city={Lyngby}, country={Denmark}}
\affiliation[inst3]{organization={Université Grenoble Alpes, CNRS, Grenoble INP, 3SR},city={38000 Grenoble}, country={France}}

\begin{abstract}
Bio-cementation uses bacterially induced calcite to bind sand grains, offering a low-carbon approach to soil stabilization. However, the 3D morphology, orientation texture, and internal strain states of individual calcite bonds remain insufficiently characterized. Here, we combine computed micro-tomography, 3D X-ray Diffraction (3DXRD), and Dark-Field X-ray Microscopy (DFXM) to nondestructively characterize grain morphology, crystallographic orientation, and both type II (intergranular) and type III (intragranular) elastic strains in calcite formed at sand–sand contacts during bio-cementation. Tomography establishes the sample morphology and the cemented contact architecture; 3DXRD provides grain-averaged orientation and strain states; and DFXM resolves sub-grain misorientations and localized strain concentrations generated during growth with 100 nm resolution. The combined results show that calcite precipitation through bio-cementation produces anisotropic internal strain and distinct sub-domain structures that can influence bond integrity and load transfer at the macroscopic scale.
\end{abstract}


\end{frontmatter}

\section{Introduction}

Cementitious materials form the cornerstone of modern construction, providing the essential binding phase that imparts cohesion, strength, and durability to aggregates in concrete and related composites. Their macroscopic performance is intrinsically linked to microstructural characteristics --- namely the arrangement, morphology, and crystallography of hydration or precipitation products. Over the past decades, extensive research has elucidated the chemistry and microstructure of hydrated Portland cement phases such as calcium–silicate–hydrate (C–S–H), portlandite, and ettringite, highlighting how their evolution governs mechanical and transport properties of cement-based materials \cite{Hu2019, Mazaheripour2018, Richardson2008}. Advanced characterization techniques, including X-ray diffraction \cite{Morales-Cantero2024}, electron microscopy \cite{stutzman2004scanning}, and synchrotron-based imaging \cite{Gallucci2007}, have been instrumental in linking microstructural parameters of cementitious materials to their bulk behavior.\\

While conventional cement remains the backbone of modern infrastructure, its production has a substantial environmental footprint, particularly due to CO$_2$ emissions from clinker manufacture and limestone calcination. These challenges have driven growing interest in sustainable alternatives. Bio-cementation, particularly in the form of Microbially Induced Calcite Precipitation (MICP), has emerged as a promising low-carbon approach for stabilizing granular soils through the precipitation of calcium carbonate (CaCO$_3$) \cite{Ferris1996, StocksFischer1999, Ehrlich1999, Urzai1999, Castanier2000}. In this process, microorganisms --- primarily Sporosarcina pasteurii --- are injected into the soil along with a cementitious solution (a urea- and calcium-rich solution). The bacteria locally alter the chemistry of the solution by hydrolyzing the urea, which raises the pH and triggers the precipitation of calcium carbonate in the form of calcite \cite{ Porter2021}. This calcite binds the soil particles together, enhancing soil stability. Such systems provide a pathway toward lower-carbon, more sustainable materials while retaining the fundamental binding mechanisms that define cementitious behavior.\\

Despite advances in the macroscopic characterization of bio-cemented materials, the microstructure of biogenic calcite itself remains poorly understood. Existing studies have primarily relied on traditional imaging techniques such as scanning electron microscopy (SEM) and X-ray tomography \cite{Dadda2017, Terzis2018, Dadda2018}. In contrast to the conventional techniques, \cite{Clark2015} employed Bragg Coherent Diffraction Imaging (BCDI) to investigate the dislocation network in calcite during growth and dissolution, using an enzyme-hydrolysis system rather than bacterial activity. More recently, Scanning 3D X-ray Diffraction (s-3DXRD) was combined with tomography to study interfaces between calcite and quartz (sand) grains, revealing that biogenic calcite is polycrystalline and displays preferential growth orientations \cite{LaBella2025}. However, the techniques used present some limitations. While BCDI provides exquisite nanoscale resolution of strain fields within individual crystals, it is inherently limited to isolated microcrystals, making it challenging to study bulk or polycrystalline materials. Scanning 3DXRD maps grain-resolved orientation and strain in bulk samples by rastering a focused X-ray beam. While the spatial resolution is determined by the beam size, the point-by-point scanning nature of the technique leads to rapidly increasing acquisition times as the beam is focused to smaller dimensions. As a result, achieving nanometer-scale resolution over extended sample volumes is challenging in practice, given the limited beamtime typically available. Both methods can therefore be impractical for investigating complex, heterogeneous materials with numerous embedded or interacting grains, such as calcite obtained from bio-cementation.

In this work, we employ a multi-modal synchrotron-based approach using 3D X-ray Diffraction (3DXRD) and Dark-Field X-ray Microscopy (DFXM) to investigate the microstructure of bio-precipitated calcite. 3D X-ray Diffraction (3DXRD) and Dark-Field X-ray Microscopy (DFXM) provide complementary capabilities for investigating the complex microstructure of biocemented calcite. While scanning 3DXRD (s-3DXRD) uses a focused beam, the 3DXRD technique uses a large beam. It allows to characterize a 3D volume faster then s-3DXRD. On the other hand, its spatial resolution is worse. 3DXRD \cite{poulsen2004three} is particularly effective for measuring Type 2 strains (long-range, grain-averaged lattice distortions across a crystal) and for determining the orientations and positions of multiple grains, which is crucial for understanding the heterogeneous strain environment created by microbial mineralization. DFXM \cite{yildirim2020probing, Simons2015, Poulsen2017}, with its higher spatial resolution (around \qty{100}{\nano\metre}), can resolve Type III strains (orientation variations between subgrains or mosaic blocks within a crystal). Thus, by combining 3DXRD and DFXM, it is possible to achieve a multi-scale, in-situ characterization of biocemented calcite, capturing both the average strain behavior across the bulk and the internal, subgrain-scale strain distributions induced by bio-cementation processes. Together, these techniques reveal how crystal growth within a sand matrix produces local misorientations and strain heterogeneities that may influence the mechanical behavior of the cemented material. This multi modal approach is particularly advantageous for studying bulk or embedded crystals, providing access to multiple grains and subgrain structures without the need for isolated microcrystals or long scanning acquisitions.

The following sections describe the experimental methods, including sample preparation and data acquisition. Results are then presented and analyzed, followed by a discussion of the key findings and limitations. The paper concludes with perspectives for future research.

\section{Materials and methods}
\subsection{Sample preparation}
The silica sand that is used in this study is denoted as S30, with round particles of diameters ranging between \qtyrange{400}{900}{\micro\metre}. This sand was also used for the experiments in \cite{Sarkis2023} and \cite{Sarkis2025}.\\
The measurements in this study were performed on two different samples. Sample 1 (S1) is a single sand grain with calcite crystals on its surface, that has been extracted from a bulk sample with a mass fraction of calcite of around \qty{6}{\percent}. The second sample (S2) consists of one cemented contact between two sand grains that was extracted from a bulk sample with a mass fraction of calcite of around \qty{15}{\percent}. For both samples, the mass fractions of calcite in the bulk samples from which they were extracted were measured using Thermogravimetric Analysis (TGA). The cementation protocol applied is the same as the one presented in \citep{Sarkis2023}, and is described below:
\begin{enumerate}
    \item The sample is first saturated with commercial water (Cristaline-Ste Cécile)
    \item 2 P.V. (Pore Volume) of the bacterial solution is injected, which consists of \qty{1}{\gram\per\litre} of Bacteria of \textit{Sporosarcina Pasteurii} and \qty{3}{\gram\per\litre} of sodium chloride (\ce{NaCl}).
    \item After a waiting time of around 4 hours, 1 P.V. of the cementing solution is injected, which consists of \qty{1.4}{\mole\per\litre} of urea and \qty{1.4}{\mole\per\litre} of calcium chloride (\ce{CaCl2}).
    \item After 2 hours, the previous \ce{CaCl2} injection is repeated.
    \item Steps 2-4 represent 1 cementation cycle. Once all the cycles are done, the sample is again flushed with water to remove any salt residues.
\end{enumerate}

After their extraction, both S1 and S2 were glued to the tips of glass capillaries using optical glue. Afterwards, three types of measurements were performed on ID03 beamline at the ESRF \cite{Isern2025}. It should be noted that in this study, the old goniometer was used, as described in \cite{Kutsal2019}. These measurements are described in the two following sections. The experimental data can be found at the corresponding dois: IH-MA-527 \url{https://doi.org/10.15151/ESRF-ES-1824795310}, IH-ES-143 \url{https://doi.org/10.15151/ESRF-ES-1896360695} and IH-ES-150 
\url{https://doi.org/10.15151/ESRF-ES-1949226012}.

\subsection{Computed X-ray tomography}
Absorption-based X-ray tomography was performed on both samples S1 and S2. The data was collected on a pco.Edge 5.5 camera from Optique Peter with a compact white beam microscope head. Different objectives can be mounted to obtain different magnifications. For S1, the data was collected at a photon energy of \qty{19}{\kilo\electronvolt} with a sample-detector distance of \qty{16.1}{\milli\metre}, and a pixel size of \qty{1.33}{\micro\metre}. For S2, the data was collected at \qty{17}{\kilo\electronvolt}, with a sample-detector distance of \qty{55}{\milli\metre}, an energy of \qty{17}{\kilo\electronvolt}, and a pixel size of \qty{1.33}{\micro\metre}. The collected data was reconstructed using the software TomWER developped at the ESRF \cite{h_payno_2022_6077027, paleo_2024_11104029}. The images were segmented using a combination of Fast Random Forest machine algorithm using Weka trainable segmentation \cite{Arganda-Carreras2017} and Python scripts, as described in \cite{Sarkis2023} and \cite{Sarkis2025}. 

\subsection{3DXRD measurements}
\begin{figure}[ht]
\center
\includegraphics[scale=0.9]{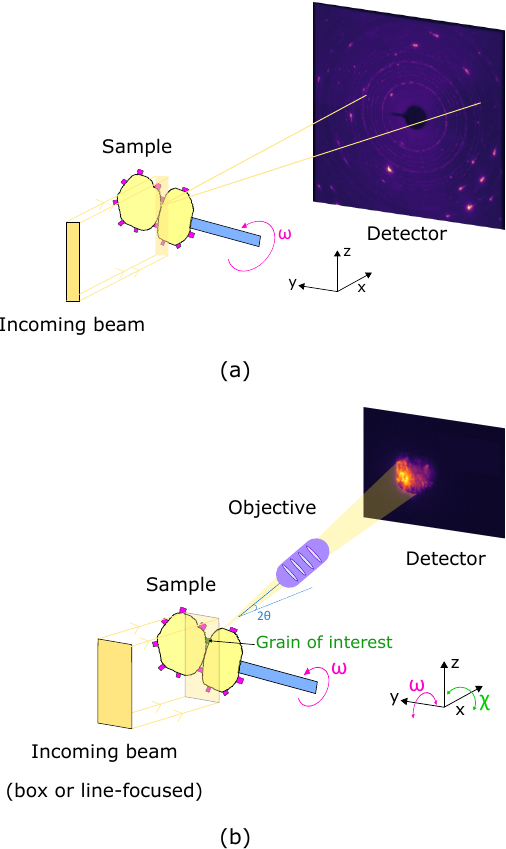}
\caption{(a) Schematics of 3DXRD setup. A box beam illuminates the region of interest in the sample. A 2D detector then collects diffraction coming from individual grains as a function of the rotation angle $\omega$. The peaks are then indexed, their positions and relative size and strain are measured, and (b) Schematics of the DFXM setup. A parallel or line beam illuminated the region of interest in the sample. Images of one diffracted lattice plane for one grain of interest (highlighted in green) are collected in function of the tilt motors $\omega$ and $\chi$ for mosaicity measurements, as well as the 2$\theta$ angle for strain measurements. In both figures, the sand is represented in yellow and the calcite in purple.}
\label{fig:3dxrddfxmsetup}
\end{figure}

3DXRD is a far-field 3D X-ray diffraction technique that consists of illuminating the sample with a box beam while rotating the sample transverse to the beam direction. The diffraction pattern is then collected, and after calibrating the detector, the phases present in the sample can be indexed \cite{Ball2024, Poulsen2021}. The grain orientations, center of mass positions in the laboratory frame, as well as their relative size and full strain tensor can be reconstructed. The setup is illustrated in \autoref{fig:3dxrddfxmsetup}a, where the sample is shown to be rotating around the y-axis, which is the configuration adopted in the current study. This measurement was performed at a photon energy of  \qty{48}{\kilo\electronvolt}, with a beam width of \qty{0.05}{\milli\metre} and height of \qty{1}{\milli\metre}.\\
\autoref{fig:tomo}b shows a reconstructed 3D volume of S2, where the calcite and sand phases are identified. Only the contact zone between the two sand grains was measured using 3DXRD, and is highlighted in the aforementioned figure. The data was collected on a FReLoN detector (ESRF) \cite{Isern2025} with a pixel size of \qty{47.3}{\micro\metre}.
\newline
The data was analyzed using the software ImageD11 \cite{wright_fable-3dxrdimaged11_2024}. Following a detector calibration with lanthanium hexaboride \ce{LaB6} powder, the sample-detector distance was found to be around 190.46 mm, and the refined photon energy to be \qty{48}{\kilo\electronvolt}. The indexing is performed in three steps:
\begin{enumerate}
    \item Peak segmentation: in this step, the background noise is removed, and the diffraction spots are detected. These peaks, and their centroid coordinates are recorded along with their intensity and associated $\omega$ angle.
    \item Geometry refinement: in this step, the geometry parameters, such as the sample-detector distance, detector tilts, beam center and sample center-of-rotation position are refined.
    \item Indexing: first, grains are indexed. Candidate grain orientations are identified based on observed angles between scattering vectors, computed from a small subset of all observed peaks. Then, these candidate orientations are scored against the remaining scattering vectors, keeping grains that score a sufficient number of peaks within a tolerance threshold. The final result is a list of grain orientations $U_i$ (rotation matrices that map the orthogonal grain reference frame to the laboratory reference frame) and $B_i$ matrices, which map (non-orthogonal) reciprocal space lattice vectors into the orthogonal grain reference frame.
    \item Grain refinement: with a list of $U$ and $B$ matrices identified, the identified matrices, alongside the centre-of-mass position of each grain, are refined against a larger subset of the peaks, yielding a list of refined grain orientations, B matrices (which encode the elastic lattice strains) and positions.
\end{enumerate}

\subsection{DFXM measurements}
Dark-Field X-ray microscopy is a full field imaging technique using diffraction contrast. It consists of isolating the Bragg diffracted beam of a crystalline element of interest (grain or domain) and passing it through a magnifying objective lens, located between the sample and a high-resolution imaging detector \cite{yildirim2020probing, poulsen2017x}. The image collected is then a magnified real-space image in a given lattice plane of the crystal that is diffracted. It can be seen as a 'zoom-in' on one of the grains that is indexed using 3DXRD. The DFXM setup is shown in \autoref{fig:3dxrddfxmsetup}b. \\
In the obtained image, contrast can be observed from the defects present within the grain. Different types of measurements can be performed. The first one is called a 'mosaicity' scan, which measures the shear/rotational strain in the grain, highlighting the different sub-domains and their spread in angles. For this, the sample has to be rotated around the y-axis by varying the $\omega$ angle of the goniometer, and around the x-axis by varying the $\chi$ angle of the goniometer. This measurement can be coupled with a tilt of the objective along the 2$\theta$ angle to measure the d-spacing differences across the grains. These are the measurements adopted in this study, and they involve 3 motors at the same time. With this technique on ID03 beamline, the spatial and angular resolutions can reach 30-100 nm, and \qty{0.001}{\deg}. The strain resolution is of around 10$^{-4}$.\\
As detailed in \cite{Poulsen2021}, with DFXM the deformation gradient tensor $\mathbf{F}$ is measured, which links the deformed state of the crystal to its undeformed state, such that:
$$\mathbf{r} = \mathbf{F} \cdot \mathbf{r}_0$$
where $\mathbf{r}$ is a position in the deformed system, and $\mathbf{r}_{0}$ a position in the undeformed system. $\mathbf{F}$ can be decomposed as:
$$\mathbf{F} = \mathbf{R} \cdot \mathbf{U} = \mathbf{V} \cdot \mathbf{R} $$
where $\mathbf{R}$ is the orthogonal rotation tensor, $\mathbf{U}$ the symmetric right stretch tensor, and $\mathbf{V}$ the symmetric left stretch tensor. Then $\mathbf{V}$ can be expressed as:
$$\mathbf{V} = \mathbf{R} \cdot \mathbf{U} \cdot \mathbf{R}^{-1}$$
Then, for small deformations, $\mathbf{F}$ can be approximated as a sum:
$$\mathbf{F} = \epsilon + \psi + \mathbf{I}$$
where $\mathrm{\epsilon}$ is the strain tensor, and $\mathrm{\psi}$ the rotation tensor in the crystal.
With the above described measurement, three components of the deformation gradient tensor $\mathbf{F}$ are measured. Two are shear/rotational, which can be obtained through the mosaicity measurement, and the third is an axial component, that can be obtained through scanning 2$\theta$.
\newline
The data analysis is performed using a dedicated software, \textit{darfix} \cite{Ferrer2023}. After several pre-processing steps, mainly consisting of determining the ROI and denoising the images, the intensity of each pixel in the raw images is processed as function of the motor positions. Motor scans thus yield maps of the moments, where each pixel corresponds, e.g., to the center-of-mass (COM, first moment) or variance ($\sigma^2$, second moment) computed over the corresponding motor. Alternatively, the intensity of a given pixel as function of motor positions can be fitted to a Gaussian, using:

$$\mathrm{FWHM} = 2\sqrt{2\ln 2}\,\sigma \approx 2.355\,\sigma$$

FWHM values extracted from the DFXM data reflect local lattice quality, with smaller widths signifying more ordered crystal regions.
\newline
The aforementioned measurements can be done in two modes: box or line-focused beam. In the first mode, the beam is illuminating the entire grain, and therefore the images collected represent a projection of the grain, where information along the depth of the grain is integrated along the line of sight, i.e. the scattered beam direction. In the second mode, the beam is focused before the sample with a set of \num{58} 1D lenses, that are around \qty{720}{\milli\metre} away from the sample at a photon energy \qty{17}{\kilo\electronvolt}, and around \qty{900}{\milli\metre} away from the sample at a photon energy of \qty{19}{\kilo\electronvolt}. This yields a beam height of around \qty{500}{\nano\metre}. This allows to 'slice' through the grain to gain access to information along its depth.
\newline

For sample S1, one calcite grain was imaged at the surface - denoted as G1 at a photon  energy of \qty{19}{\kilo\electronvolt}, while for sample S2, the focus area was also the contact area, where two other calcite grains were imaged as well - denoted as G2 and G3, at a photon energy of \qty{17}{\kilo\electronvolt}. For G1 and G2, the lattice plane \hkl(10-14) was imaged, and for G3 the lattice plane \hkl(20-22).

\section{Results and Analysis}

\subsection{Computed X-ray tomography and 3DXRD measurement results}
\begin{figure}[htbp]
\center
\includegraphics[scale=0.85]{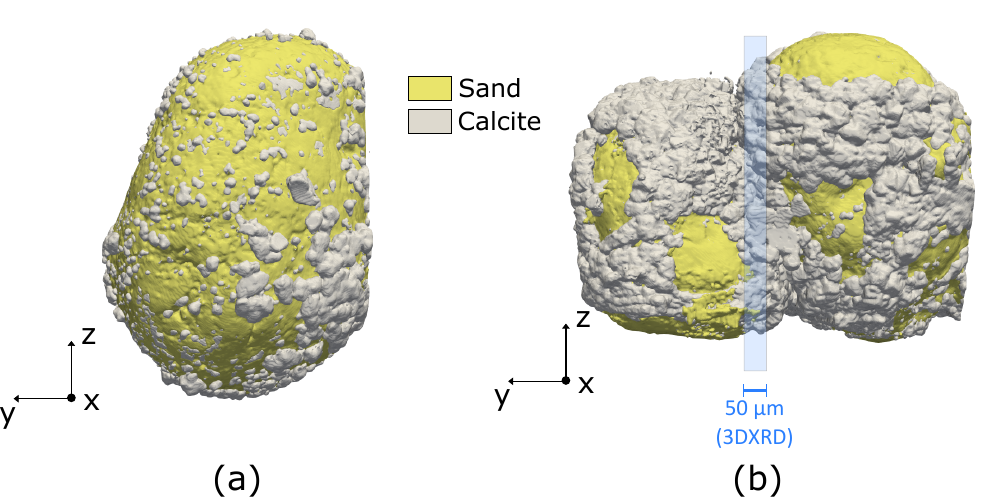}
\caption{3D rendering of the volumes acquired with tomography of: (a) S1, and (b) S2, showing the concentrated beam at the contact for the 3DXRD measurement.}
\label{fig:tomo}
\end{figure}

\autoref{fig:tomo}a and b show the reconstructed 3D volumes of samples S1 and S2 respecively. The sand is shown in yellow and the calcite crystals are shown in white. No direct quantitative analysis was performed on the datasets, which were rather used to get an overview of the sample and prepare for the 3DXRD and DFXM measurements that follow.

\begin{figure}[htbp]
\center
\includegraphics[scale=0.7]{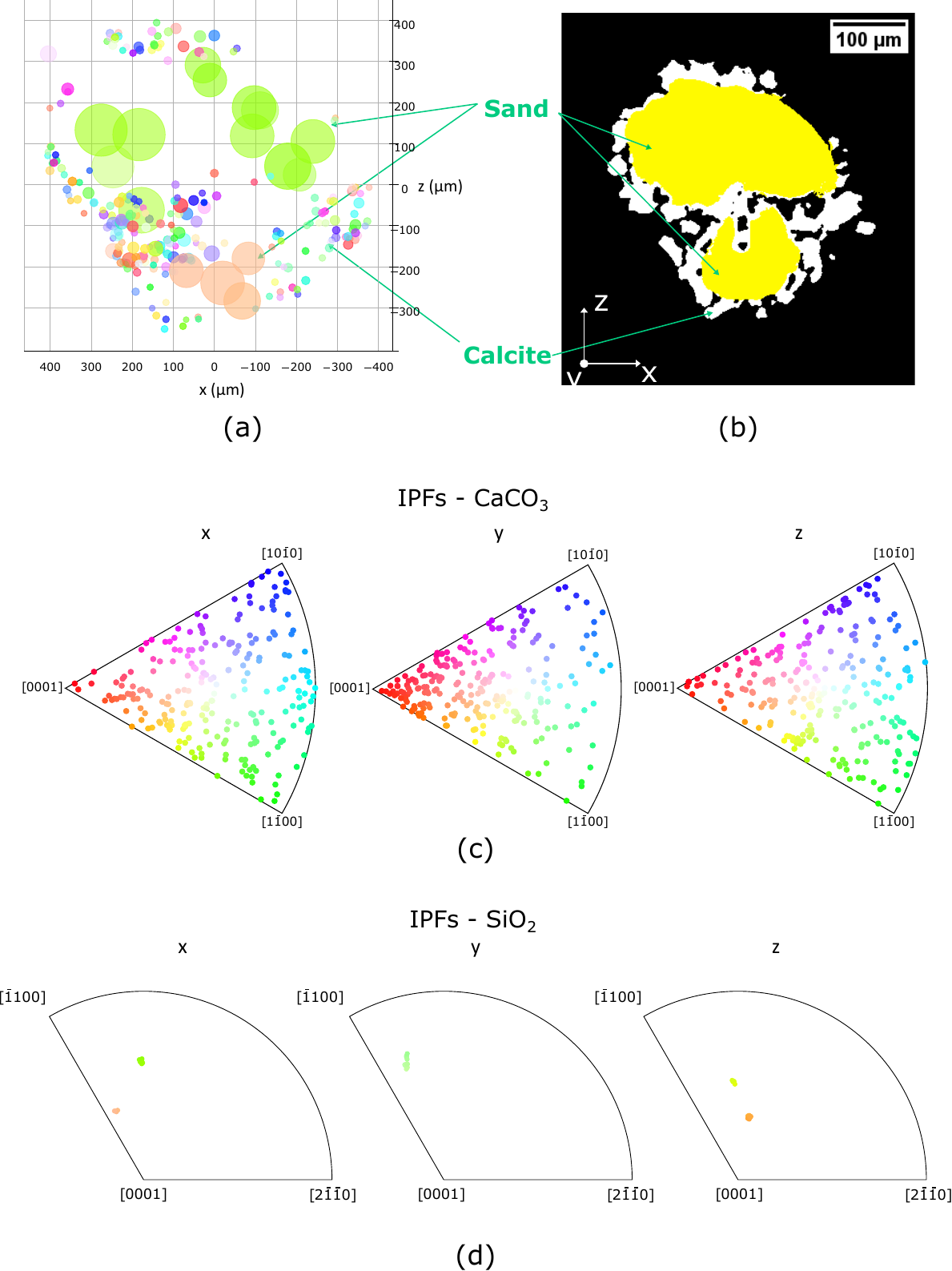}
\caption{(a) 3D plot showing the indexed sand and calcite grains from 3DXRD (colored by IPF-x). Axes values are centered around the center of the incoming beam, (b) the segmented mid-layer tomography slice from the illuminated volume for 3DXRD (white: calcite, yellow: sand), (c) the IPFs for the calcite crystals along the laboratory x, y and z directions, and (d) the IPFs for the quartz crystals along the laboratory x, y and z directions.}
\label{fig:ipfs}
\end{figure}

3DXRD allows for the reconstruction of the positions and orientation as well as the full elastic strain tensor of the grains. The grain sizes obtained are only relative, and the radius of each grain is computed by estimating the grain size from the summed intensity of all the peaks associated to it \cite{oddershede2010determining}.\\
\autoref{fig:ipfs}a shows the indexed grains from 3DXRD, both sand and calcite. 236 calcite and 16 quartz crystals (12 for one sand grain, 6 for the other) were indexed from the illuminated volume. This 3DXRD reconstruction matches with the reconstructed slice in \autoref{fig:ipfs}b from tomography, which corresponds to the mid-layer in the \qty{50}{\micro\metre} volume illuminated during the 3DXRD measurement, as shown in \autoref{fig:tomo}b. The two reconstructed quartz grains are actually shown as two clusters of particles, mainly due to the presence of microcracks in the sand grains, resulting in the detection of many grains with similar orientations. The IPFs (Inverse Pole Figures) for the quartz grains are shown in \autoref{fig:ipfs}d. The two clusters happen to have very similar orientations along the y-axis of the laboratory. On this map, the colors represent the orientations of the grain in the laboratory coordinates, and the sizes represent the intensity of the diffracted spot, which makes it rather a relative quantity. \\

\autoref{fig:ipfs}c shows the inverse pole figures (IPFs) along the x, y and z laboratory coordinates for the calcite crystals. We observe that a preferential growth orientation is present: the c-axis of the calcite crystals is mostly parallel to the y direction. As shown in \autoref{fig:tomo}b, the y-axis is roughly perpendicular to the contact between the two sand grains. This preferential growth orientation is also observed in \cite{LaBella2025} from scanning 3DXRD data on layers at the contact between the two sand grains. As stated in the aforementioned study, it is highly unlikely that this texture is the result of any epitaxial growth of calcite on quartz, given their different chemical and crystallographic properties.\\

\begin{figure}[htbp]
\center
\includegraphics[scale=0.8]{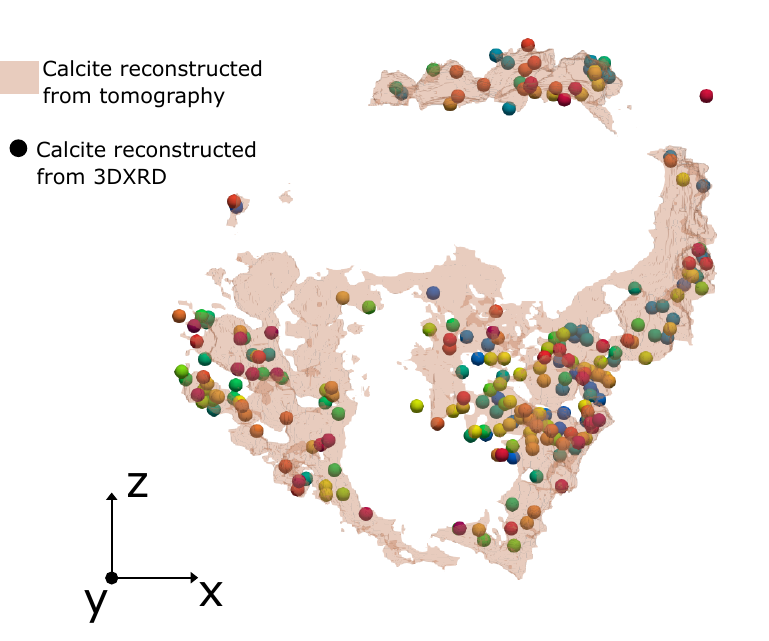}
\caption{Overlay of the reconstructed calcite crystals from 3DXRD (colored by grain ID) and the reconstructed illuminated calcite volume from tomography. Refer to \autoref{sec:3dxrdpbs} for the explanation behind missing regions in the 3DXRD reconstruction.}
\label{fig:tomo-3dxrd}
\end{figure}

From \autoref{fig:tomo-3dxrd}, by combining the indexed grains map issued from 3DXRD and the tomography, it can be observed that the calcite crystals are stacked on top of each other, especially at the borders around the sand grains. This indicates that from cycle to cycle, the calcite crystals are deposited on top of those grown during the previous cycle, as the new batch of bacteria is injected and new nucleation sites are created. \\

\begin{figure}[htbp]
\center
\includegraphics[scale=0.9]{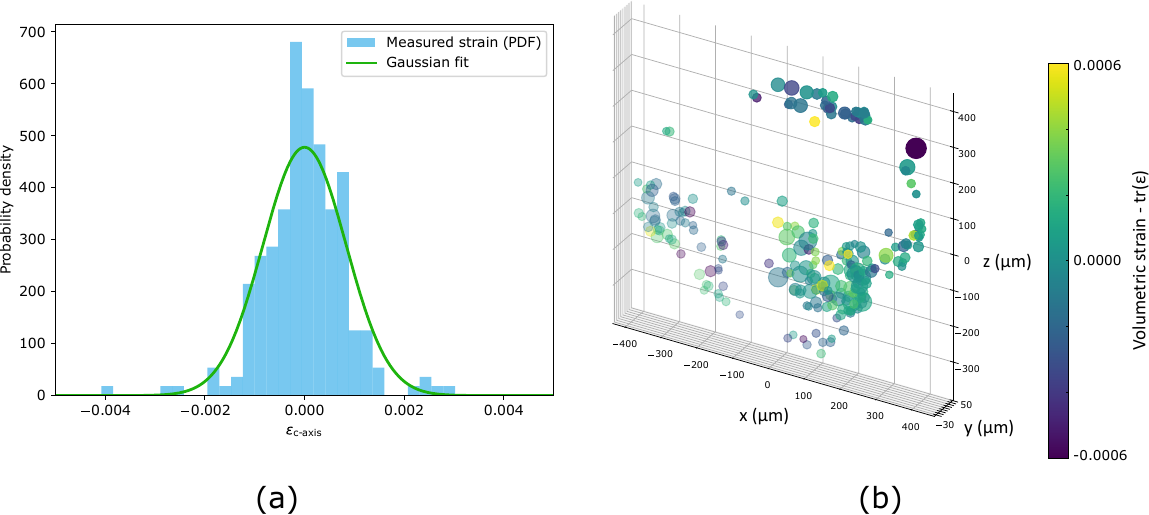}
\caption{(a) Histogram of the relative strain of the calcite crystals along the c-axis (in crystal frame), and (b) 3D plot showing the indexed calcite grains from 3DXRD, colored by the volumetric strain. Image shows with display range between -6$\times 10^{-4}$ and 6$\times 10^{-4}$ for visualization purposes. The raw volumetric strain data ranged between -1.6$\times 10^{-3}$ and 3$\times 10^{-3}$.}
\label{fig:epscaxi}
\end{figure}

The relative strain along the c-axis in the crystal frame is plotted as a probability density histogram in \autoref{fig:epscaxi}a. Most of the calcite crystals show a small strain value along their c-axis. However, as it can be observed from the histogram, significant strain is present, which is in the order of \num{e-3}, both in compression and tension. This histogram was fitted with a gaussian function, with a standard deviation of 8.34 $\times$ $10^{-4}$. The mean value used for the fit was -2.32 $\times$ $10^{-7}$. It should be noted that this value is very small and below the noise floor for strain with 3DXRD, which is in the order of $10^{-4}$. Since these crystals are protected by the contact (i.e. they cannot be detached from the surface of the quartz grains), these values of strain are mainly due to the growth of the calcite crystals.

\autoref{fig:epscaxi}b shows the volumetric strain (tr($\epsilon$)) measured in the laboratory coordinates. The mean volumetric strain measured is \num{1.3e-4} $\pm$ \num{6.2e-4}, indicating that on average, the calcite crystals have low volumetric deformation.

\autoref{tab:crystalparam} shows the mean computed lattice parameters of both the calcite and quartz phases, with respect to which the strains were computed in the case of the calcite crystals. These values are in accordance with the calcite parameters obtained from refined laboratory powder diffraction data on the same material in \cite{LaBella2025}.

\begin{table}[htbp]
\centering
\caption{Mean crystal structure parameters for calcite and quartz (XRPD: X-ray Powder Diffraction).}
\label{tab:crystalparam}
\begin{tabular}{lcccc}
\toprule
\textbf{Phase} & \textbf{Technique} & \textbf{\(a = b\) (\AA)} & \textbf{\(c\) (\AA)} & \textbf{Space Group} \\
\midrule
Calcite (CaCO$_3$) & 3DXRD & 4.9911 & 17.0947 & 167 \\
S\_lab\tablefootnote{Data from \cite{LaBella2025}.} (CaCO$_3$) & Lab XRPD & 4.9922 & 17.089 & 167 \\
Quartz (SiO$_2$) & 3DXRD & 4.9133 & 5.4039 & 152 \\
\bottomrule
\end{tabular}
\end{table}

\subsection{DFXM results}
\begin{figure}[htbp]
\center
\includegraphics[scale=0.8]{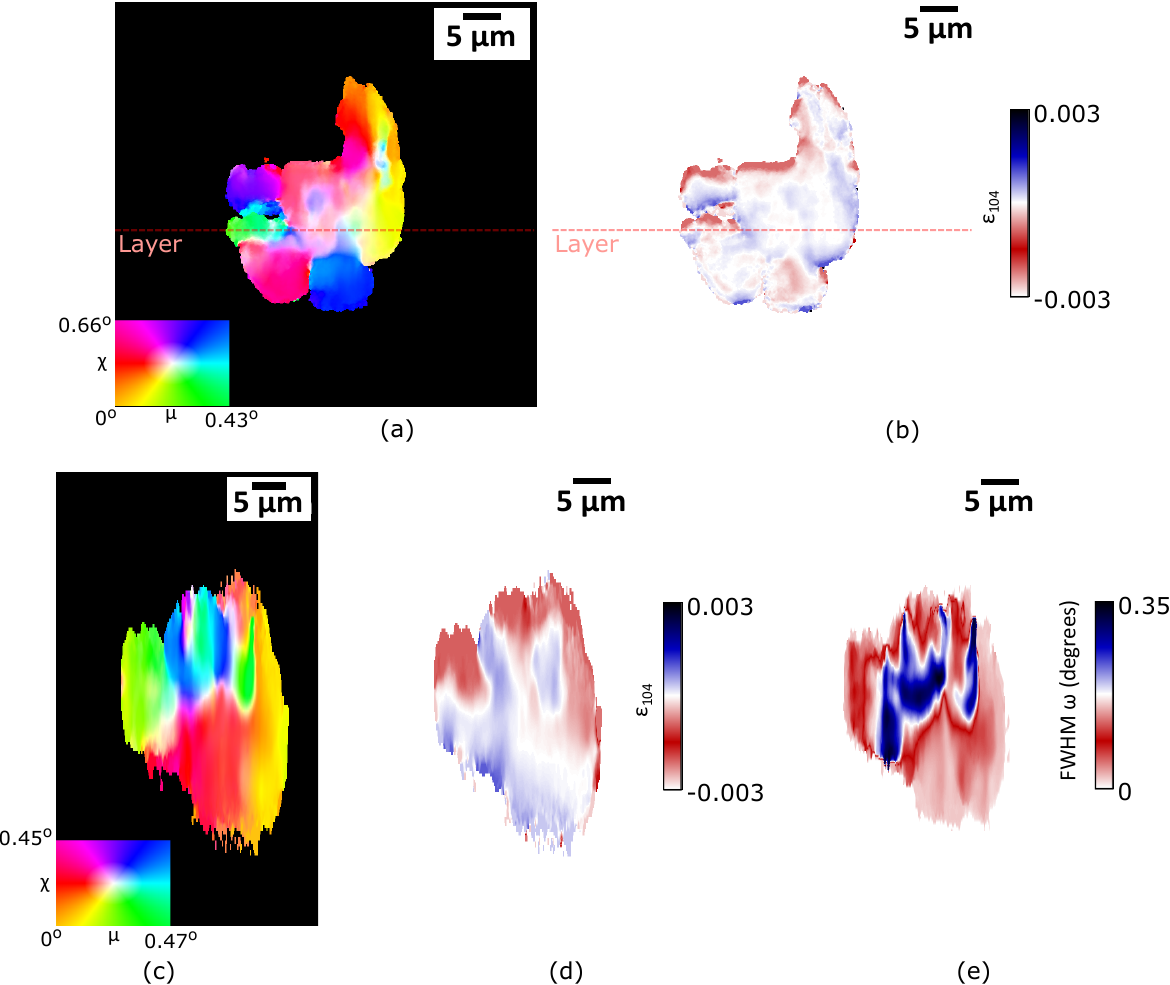}
\caption{For G1, for lattice plane  \hkl(10-14): (a) Mosaicity map of a projection, (b) relative strain map projection, (c) mosacity map in a layer, (d) relative strain map in the same layer, and (e) FWHM map of the $\omega$ angle in the same layer.}
\label{fig:g1dfxm}
\end{figure}

Both samples S1 and S2 were subject to Dark-Field X-ray Microscopy (DFXM). For each sample, one calcite crystal was measured. The grain measured in sample S1 will be denoted by G1, and the one measured in sample S2 will be denoted by G2. In both cases, the lattice plane measured was \hkl(10-14) ($d = \qty{3.035}{\angstrom}$), which is also the most intense reflection for calcite. \\
\autoref{fig:g1dfxm} shows mosaicity and strain maps obtained for G1, both in projection and layer modes. \autoref{fig:g1dfxm}a shows the projected mosaicity map of G1, where it can be observed that G1 is indeed made up of many sub-domains, that are well-defined, and are misoriented with respect to each other along the height and width of the grain. \autoref{fig:g1dfxm}b shows that the strain is relatively homogeneous across the grain.\\
The beam was then focused into an intense line-beam of around \qty{500}{\nano\metre} in height at the level of the red dashed line shown in Figures \ref{fig:g1dfxm} a and b. The same measurements were performed, but by observing a layer sliced into the grain. \autoref{fig:g1dfxm}c shows the mosaicity map of this layer, in which we can also see distinctive sub-domains, that are well-defined with sharp grain boundaries that appear in white in the mosaicity plot. When compared to the relative strain plot in \autoref{fig:g1dfxm}d, it can be observed that the strain in many cases, is localised in areas of the grains where well-defined sub-domains exists. The combination of elastic and plastic strain with local misorientation suggests that this region of the calcite grain reflects a growth-adaptive response, where the crystal lattice accommodated internal and external constraints during bio-cementation. This combination shows as a signature in high FWHM values in \autoref{fig:g1dfxm}e. However, the highest values in the FWHM map are located at the sub-grain boundaries, indicating that these interfaces concentrate misorientation and plastic deformation within the crystal.

\begin{figure}[htbp]
\center
\includegraphics[scale=0.7]{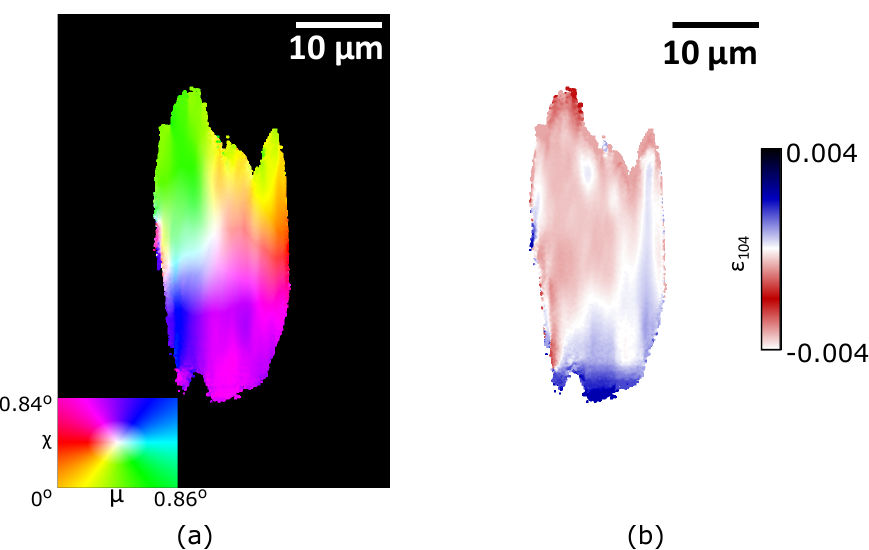}
\caption{For G2, from projection measurements on lattice plane $(1\,0\,\bar{1}\,4)$: (a) mosaicity map, and (b) relative strain map.}
\label{fig:g2proj}
\end{figure}

Equivalent DFXM measurements were conducted on grain G2, and the resulting mosaicity and strain projections are shown in \autoref{fig:g2proj}. Similar to G1, G2 displays pronounced mosaicity and the presence of multiple subdomains distributed along both the height and width of the grain. In contrast to G1, however, the relative strain map of G2 reveals higher strain magnitudes, suggesting a more heterogeneous internal lattice distortion. 

\begin{figure}[htbp]
\center
\includegraphics[scale=0.75]{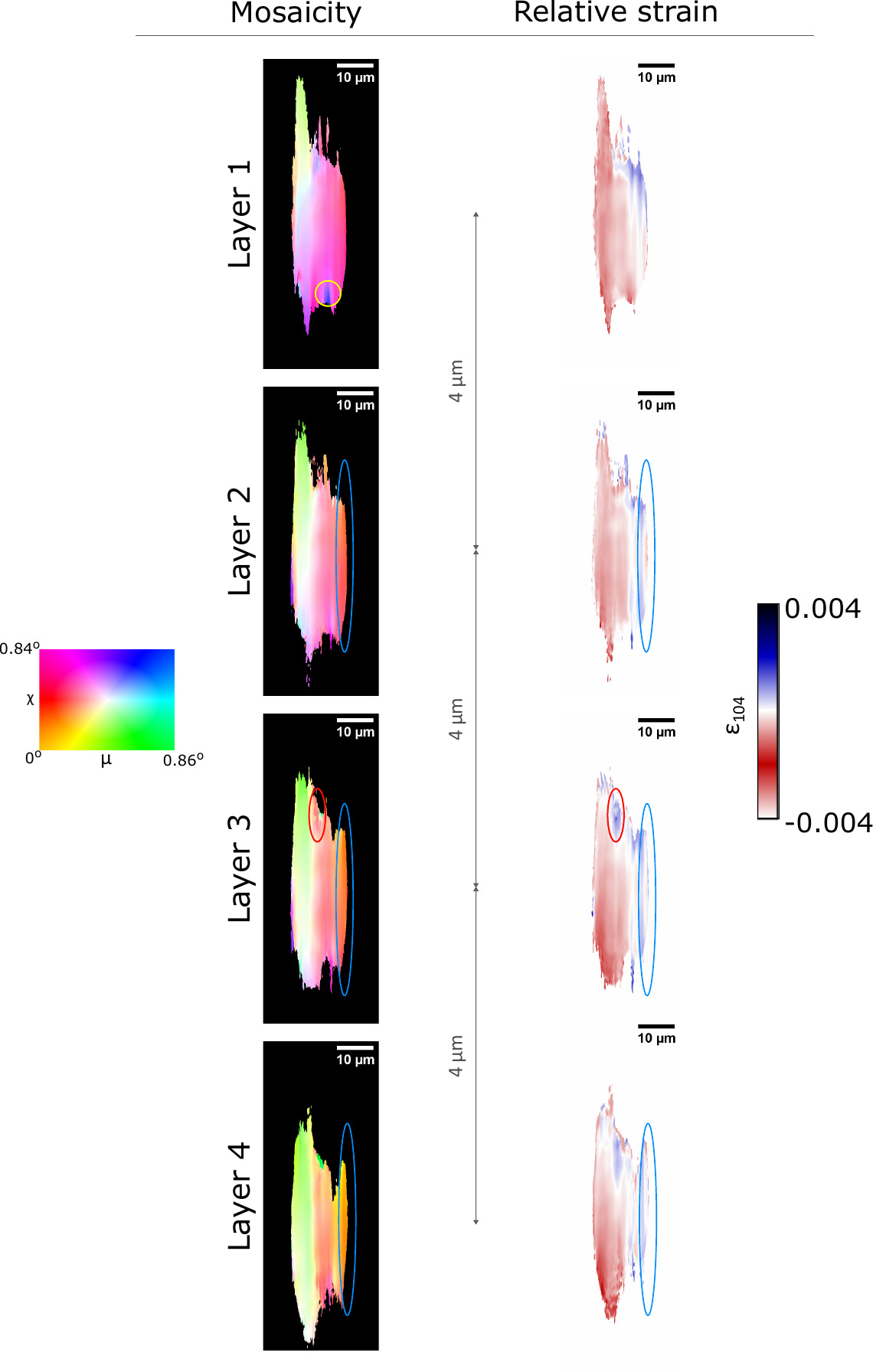}
\caption{Mosaicity and relative strain maps for different layers across G2.}
\label{fig:g2layerscans}
\end{figure}

\autoref{fig:g2layerscans} presents the mosaicity and relative strain maps of grain G2 for layers spaced \qty{4}{\micro\metre} apart along the grain height. The observed variations in mosaicity indicate a gradual change in grain orientation with height. From layer 2 onward, distinct sub-domains appear, separated by a sharp grain boundary, which suggests a twisting of the calcite crystal. As highlighted by the blue ellipses in \autoref{fig:g2layerscans}, the right-hand region of the grain exhibits a slight misorientation associated with compressive strain. This behavior implies that the grain experienced localized compression, likely imposed by either an adjacent growing calcite crystal or neighboring sand grains, leading G2 to twist in order to accommodate further growth. Similarly to G1, and as highlighted with the yellow discs, misorientation is also coupled with localised strain in many regions of the grain G2.

\begin{figure}[htbp]
\center
\includegraphics[scale=0.75]{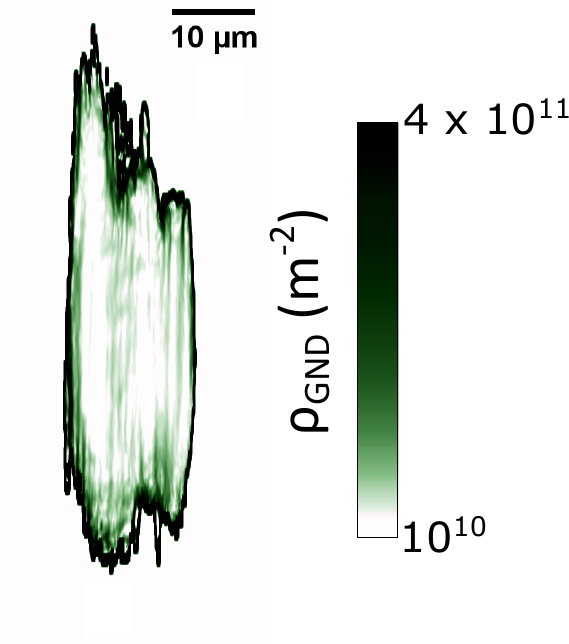}
\caption{Geometrically necessary dislocations (GND) density map of layer 4 for sample S2. Image shown with display range between $0$ and $2\times 10^{11} \,\mathrm{m}^{-2}$ for visualization purposes. Raw GND density values ranged between 0 and $7.72\times 10^{12} \,\mathrm{m}^{-2}$.}

\label{fig:s2-layer4_gnd}
\end{figure}

In \autoref{fig:g2layerscans}, layers 2, 3 and 4 show sharp sub-grain boundaries, that appear as white lines separating misoriented zones in the calcite grains in the mosaicity maps. These grain boundaries usually consist of an array of dislocations, and show in a higher contrast in mosaicity measurements in DFXM. For this, it is possible to compute the density of the geometrically necessary dislocations (GNDs), which will be denoted as $\rho_{\mathrm{GND}}$ in the following. First, the Kernel Average Misorientation (KAM) for the grain can be computed as:
\[
\mathrm{KAM}(i,j) = \frac{1}{N} \sum_{n=1}^{N} \lvert \phi(i,j) - \phi(n) \rvert 
\]
where $\phi(i,j)$ is the local misorientation at the pixel \textit{(i, j)}, $\phi(n)$ the misorientation of the $n_{\mathrm{th}}$ neighbouring pixel, and \textit{N} is the number of neighbouring pixels.
The density of the GNDs can be calculated as in \cite{Pal2026}:
$$\rho_{\mathrm{GND}} = \frac{2(\mathrm{KAM})}{b.s}$$
Where $b$ is the magnitude of the Burger's vector, and $s$ is the resolution step, which is around \qty{0.3}{\nano\metre} in the case of this study. Here, $b$ is assumed to be equal to \qty{0.8}{\nano\metre}, corresponding to the a-type slip system that is the most occurring in calcite crystals \cite{Ihli2016}. \autoref{fig:s2-layer4_gnd} shows the map of the density of the GNDs for layer 4 presented in \autoref{fig:g2layerscans}. The average value of the GND density was found to be around 2.57 $\times$ 10$^{11}$ m$^{-2}$, and reached $7.72\times 10^{12}$ m$^{-2}$ as a maximum value. 

\begin{figure}[htbp]
\center
\includegraphics[scale=0.75]{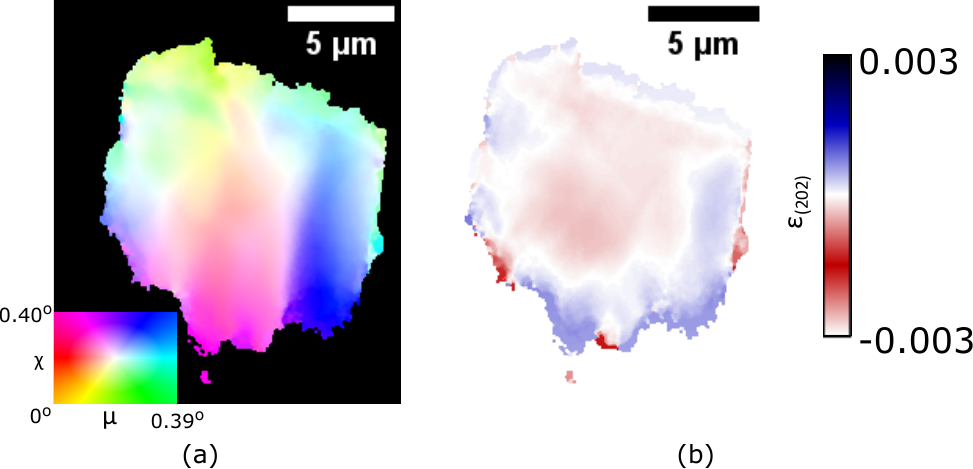}
\caption{For G3, from projection measurements on lattice plane \hkl(20-22): (a) mosaicity map, and (b) relative strain map.}
\label{fig:g3proj}
\end{figure}

A different grain, G3, was also imaged at the contact area for sample S2, this time aiming for a different lattice plane, the \hkl(20-22) plane. The objective is to check if these biogenic calcite crystals also present high mosaicity and strain in different lattice planes. As it can be observed in \autoref{fig:g3proj}a, where a mosaicity projection measurement is presented, G3 is also composed of many sub-domains showing high mosaicity, with a very similar twisting pattern across the width of the grain as the one observed in G2. The strain projection in \autoref{fig:g3proj}b also shows homogeneous strain distribution

\section{Discussion}
In this study, 3DXRD was employed as a coarse-mapping method on an illuminated volume at the contact between two sand grains in sample S2, providing the average strain tensor, crystal orientation, and spatial position of each grain. DFXM, in contrast, offered higher spatial and angular resolution, enabling detailed imaging of individual calcite crystals and revealing sub-domains as well as the relative strain distribution along specific lattice planes. The above observations show the microstructure of the bio-precipitated calcite crystals that is only due to their growth process, without any induced mechanical stress or loading. In light of the results presented above, several points merit further discussion, and are listed in the following sections.

\subsection{High grain mosaicity and local misorientation}
All grains examined with DFXM (G1 and G2 along the \hkl(10-14) reflection and G3 along the \hkl(20-22) reflection) exhibit pronounced mosaicity and internal lattice misorientation. The presence of these features in different reflections indicates that lattice distortion is not plane-specific but rather an \textit{intrinsic} characteristic of the biogenic calcite crystals. During bio-cementation, local variations in ion concentration, supersaturation, and confinement between sand grains likely lead to uneven growth rates across crystal faces. Such spatially heterogeneous precipitation conditions can promote the incorporation of small lattice rotations as the crystal accommodates non-uniform growth stresses. The resulting sub-grain structure and internal strain heterogeneity reflect a growth-induced mosaic microstructure rather than purely post-depositional deformation. From a cementitious perspective, these findings are significant because such internal heterogeneities act as stress concentrators and may influence the mechanical integrity and long-term stability of the calcite binder phase within bio-cemented sands.

\subsection{Growth-induced strain and  mechanical implications}    
As previously discussed, the strain measured by both 3DXRD and DFXM is attributed to growth-induced lattice distortions, since no external mechanical loading was applied in this study, and the samples were handled with extreme care during their preparation. The localized strain fields, misorientation, and sub-grain structures revealed by DFXM suggest the presence of potential stress concentration sites within the calcite crystals. Similar microstructural features have been reported in mechanically deformed calcite, where intragranular cracking and twinning tend to nucleate in regions of elevated internal strain and misorientation \cite{Schuster2020}. This comparison implies that the heterogeneity observed here could play a key role in controlling crack initiation and the mechanical response of biogenic calcite.

Moreover, the strain distribution is likely influenced by the growth environment at the sand–calcite interface. Each crystal nucleates on the surface of one sand grain and continues to grow until it impinges on the opposite grain, leading to potential confinement effects. Such geometrical constraints could induce compression or torsion within the growing crystal. Consequently, the “active” calcite crystals --- those bridging two sand grains --- may exhibit higher strain values than “inactive” crystals that remain attached to a single grain. This interpretation is consistent with the relatively high standard deviation observed for the strain along the c-axis (\autoref{fig:epscaxi}), which could reflect variability between these two crystal populations.

\subsection{Presence of significant volumetric strain}
As presented in the results section, large values of volumetric strains were measured in the calcite crystals from 3DXRD, reaching around -1.6$\times 10^{-3}$ in compression and 3$\times 10^{-3}$ in tension. This deformation can be coming from:
\begin{enumerate}
    \item Presence of organic matter: \cite{Perito2018} and \cite{LaBella2025} found clear evidence of the presence of an organic matrix in bio-precipitated calcite, causing a deformation of the unit cell using X-ray Powder Diffraction (XRPD). Both studies found a large \textit{c/a} ratio compared to that obtained from the reference unit-cell parameters of the calcite crystal. After being exposed to high temperatures, this anisotropic strain disappeared. Indicating that the bacteria left traces of organic molecules that were incorporated into the calcite unit-cell, and disappeared at high temperatures.
    \item Impurities: the sand surface can present some chemical impurities, containing elements such as Mg, Mn and Fe \cite{Khwaja2015}. The \ce{Ca^2+} cations in the calcite unit cell can be replaced by the \ce{Mn^2+}, \ce{Fe^2+} or \ce{Mg^2+} cations \cite{Almqvist2010}, resulting in volumetric changes of the calcite unit cell.
\end{enumerate}

\subsection{Effect of the calcite content}
Even though the two samples S1 and S2 contain different mass fractions of calcite, the DFXM measurements revealed similar microstructures for both samples. The comparison between the calcite grains G1, G2 and G3 based on this difference is not straightforward. Calcite precipitates in a heterogeneous way in the bulk, which means that the samples S1 and S2 cannot be assumed to be representative themselves in terms of calcite content. 

In addition, in the case of S2, the measurements were done at the contact, so the relative strain as well as the mosaicity are due to the growth of the crystals. However, it is important to note that since S1 presents a single sand grain, it is possible that the measured grain G1 was detached from the surface of another sand grain that was initially present in the bulk. Yet, both grains, S1 and S2 showed high mosaicity and similar strain values.

   
\subsection{Comparison with other types of biogenic calcite}
Biogenic calcite exists in the natural environment under many forms, including shells, such as that of molluscan shells. In MICP, microorganisms control the solution chemistry that triggers precipitation, but the subsequent crystal growth is abiotic, occurring without biological templating. This contrasts with strongly biologically controlled biominerals such as mollusc-shell calcite, where organic matrices regulate orientation, strain, and defect formation. Therefore, given that the precipitation mechanism is different, the comparison can only be made qualitatively. \cite{Schoeppler2022} observed the biogenic calcite in the \textit{P. nobilis} shells using DFXM. They found that the calcite crystals in these shells show high lattice distortion and local misorientation, which is aligning with the results of this study. In addition, higher concentrations of Mg led to a substitution with the Ca atoms in the calcite cell, which caused shrinkage and anisotropic distortion in the unit cell. This suggests that such internal heterogeneity could be intrinsic to calcite, whether it is biologically mediated (as in shells) or microbially induced (as in our MICP systems). On the other hand, \cite{deFrutos2023} observed using TEM (Transmission Electron Microscopy) that biogenic calcite of \textit{A. psittacus} presents both crystalline and amorphous calcite (vaterite). The crystalline phase forms sorts of lumps that are surrounded by the amorphous phase. This aspect is not observed in microbially-induced calcite, since only the calcite phase is observed.

\subsection{Technical challenges of 3DXRD}
\label{sec:3dxrdpbs}
\begin{figure}[htbp]
\center
\includegraphics[scale=0.75]{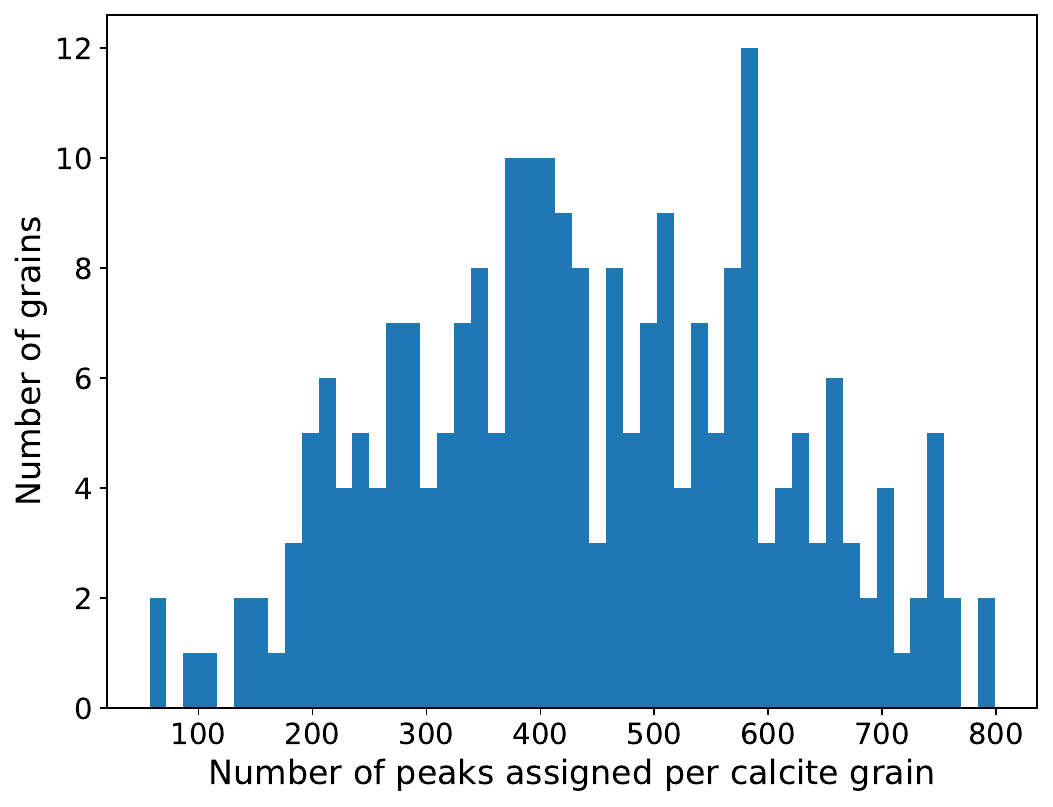}
\caption{Histogram showing the number of peaks found for the indexed calcite crystals from 3DXRD.}
\label{fig:npks}
\end{figure}

As observed in \autoref{fig:tomo-3dxrd}, some calcite grains captured in the tomography data have not been successfully indexed with 3DXRD. The reason behind this lies in the sample geometry: each sand particle has a diameter of \qty{600}{\micro\metre} on average, and is made of one or many quartz single crystals. On the other hand, bio-precipitated calcite crystal sizes range between \qtyrange{10}{50}{\micro\metre}. Since the intensity of the diffracted beam is directly proportional to the volume of the diffracting object when using a box-like beam, the diffracted spots coming from the quartz were much more intense than those coming from the calcite. Due to the limited dynamic range of the detector, this rendered the data acquisition and analysis extremely challenging -- even with a \qty{50}{\micro\metre} wide box-beam, the exposure time needed to clearly view the calcite diffraction spots caused detector saturation from the quartz peaks. These saturation events manifest as vertical streaks that can hide the calcite peaks. Reducing the exposure time to entirely avoid detector saturation caused the calcite diffraction spots to be too close to the noise floor of the detector. Even though a trade-off was found, some saturation still occurred, and some calcite diffraction peaks remained unsegmented. For this reason, the tolerances on the minimum number of peaks that a calcite crystal should have in order to be accepted were lowered. \autoref{fig:npks} shows a histogram of the number of peaks for the indexed calcite crystals from 3DXRD: 3 grains in total have less than 100 diffracted peaks. This low peak number may have induced some variance in the reconstructed grain positions or strains. For instance, the indexed calcite grain at the top right of \autoref{fig:tomo-3dxrd} has 59 diffracted peaks and does not have a match from the tomography volume, because its position might have wrongfully computed.
To overcome these aforementioned challenges, other grain-mapping techniques that use a point-like beam, thereby reducing the beam-sample interaction volume, such as scanning 3D X-ray diffraction (s-3DXRD) may be more suitable for a similar material.\\

\section{Conclusion and perspectives}
This work demonstrates that multiscale and multimodal X-ray characterization can provide unprecedented insight into the microstructural and crystallographic properties of bio-precipitated calcite in bio-cemented sand. By resolving strain gradients, lattice misorientations, and sub-grain boundaries at the single-crystal level, we reveal the fundamental heterogeneities that may govern stress accommodation within the calcite binder. Although direct mechanical testing was not performed, these observations establish a mechanistic link between biomineral formation processes and cementitious performance, positioning bio-cementation as a structurally and functionally relevant alternative to conventional binders. Our findings provide a foundational step toward designing low-carbon cementitious materials with predictable microstructure–function relationships.

The methods used were complementary, demonstrating that information beyond absorption imaging (i.e., tomography) is essential to fully understand the microstructure of calcite crystals, and that coarse-grain mapping and local diffraction-contrast imaging together reveal distinct yet consistent aspects of their internal structure. The key finding of this study are listed below:
\begin{itemize}
    \item Calcite crystals stack on top of each other due in their growth process. What appears as one large crystal from tomography data, is indeed a collection of many.
    \item They showed a preferential growth orientation, since the c-axis of the crystals were in majority pointing perpendicular to the contact between the two sand grains. 
    \item Significant relative strain values were found along the c-axis. In addition, impurities present on the surface of the sand grains, as well as the bacteria themselves, may be inducing volumetric strain in the calcite crystals. These strain values and high mosaicity are mainly due to growth.
    \item The internal microstructure of bio-precipitated calcite presents high mosaicity, with very distinct sub-domains, which is common among other biogenic calcite.
    \item DFXM also revealed that the strain inside the calcite crystals shows similar relative values as the ones obtained from 3DXRD. Localised strain is often combined with local misorientation in the grain.
    \item The GND density was found to be in the order of 10$^{11}$, computed for a layer of G2.

\end{itemize}

We believe future research should progress in four directions:
\begin{enumerate}
    \item Systematic experiments are needed to quantify how parameters, such as CaCO$_3$ content, number of cementation cycles, and solution chemistry, govern the crystallography and strain state of biogenic calcite.
    \item Microstructural measurements should be integrated with mechanical testing to determine how lattice orientation, defect structures, and type II/III strain evolve under load and influence failure in microbially-induced calcite.
    \item Multimodal–multiscale synchrotron workflows should be implemented, in which DFXM is used to follow intragranular strain evolution in selected grains during  in-situ loading, followed by full-field 3DXRD to map pre and post-deformation orientations and long-range strain across the aggregate \cite{Shukla2025}. Such integrated measurements will provide mechanistic links between grain-scale processes and macroscopic mechanical performance.
    \item The microstructre of biogenic calcite studied in this paper should be compared with other types of calcite, such as the one present in tap water. Moreover, it would be interesting to study the microstructure of biogenic calcite grown on cleaned sand, i.e. with as less impurities as possible, and evaluate the volumetric strain.
\end{enumerate}

\section*{CRediT authorship contribution statement}
\textbf{Marilyn Sarkis}: Writing --- review \& editing, Writing --- original draft, Methodology, Investigation, Data curation, Formal analysis, Conceptualization. \textbf{James A. D. Ball}:  Writing --- review \& editing, Data curation, Formal analysis, Software. \textbf{Michela La Bella}:  Writing --- review \& editing. \textbf{Antoine Naillon}:  Writing --- review \& editing. \textbf{Christian Geindreau}:  Writing --- review \& editing. \textbf{Fabrice Emeriault}:  Writing --- review \& editing. \textbf{Carsten Detlefs}:  Writing --- review \& editing, Supervision, Funding acquisition, Methodology, Investigation, Formal analysis, Conceptualization. \textbf{Can Yildirim}: Writing --- review \& editing, Methodology, Investigation, Data curation, Formal analysis, Conceptualization.

\section*{Declaration of competing interests} 
The authors declare that they have no known competing financial
interests or personal relationships that could have appeared to influence
the work reported in this paper.

\section*{Acknowledgments}
The authors acknowledge the ESRF for providing beamtime (IH-MA-527, IH-ES-143, and IH-ES-150). The authors acknowledge Leslie Sapin and Annette Esnault-Filet for providing the sample S1 in this study. They also acknowledge the help of Valérie Magnin from ISTerre for the TGA measurements. The bacteria used were produced by Soletanche Bachy (contact annetteesnault@
soletanche-bachy.com).

\clearpage
\bibliographystyle{elsarticle-num}
\bibliography{biblio}

\end{document}